\documentclass[12pt]{article}
\usepackage{amsmath,amssymb}
\begin{document}
\title{Entropy of Partitions on Sequential Effect Algebras \thanks {This project is
supported by Research Fund of Kumoh National Institute of
Technology.}}

\author{{Wang Jiamei}\thanks {Department of Mathematics, Zhejiang University,
Hangzhou 310027, People's Republic of China.},\,\, {Wu Junde}\thanks
{Corresponding author: Department of Mathematics, Zhejiang
University, Hangzhou 310027, People's Republic of China. E-mail:
wjd@zju.edu.cn},\,\, {Cho Minhyung}\thanks {Department of Applied
Mathematics, Kumoh National Institute of Technology, Kyungbuk
730-701, Korea. E-mail: mignon@kumoh.ac.kr}}
\date{}
\maketitle

{\bf Abstract} {\small By using the sequential effect algebra
theory, we establish the partitions and refinements of quantum
logics and study their entropies.}

\vskip 0.1 in



\vskip0.2in

 \noindent {\bf 1.  Introduction}

\vskip0.2in

\noindent {\it Quantum entropy} or {\it Von Neumann entropy}, which
is a counterpart of the classical {\it Shannon entropy}, is an
important subject in quantum information theory ([1]).  In order to
study the entropy of {\it partition} of {\it quantum logics}, in
[2], the author tried to define the partitions and {\it refinements}
of quantum logics, nevertheless, his methods are only suitable for
{\it classical logics}, the essential reasons are that the classical
logics satisfy the {\it distributive law} but quantum logics do not.
In this paper, by using the {\it sequential effect algebra} theory,
we establish really effective refinement methods of quantum logics
and study their entropies.

 \vskip0.2in

 \noindent {\bf 2.  Classical logics and quantum logics}

\vskip0.2in

\noindent As we know, the classical logics can be described by the
Boolean algebras and the quantum logics can be described by the
orthomodular lattices ([2-5]). The classical probability or Shannon
entropy was based on the classical logics and quantum entropy was
established on the quantum logics ([1]). Now, we recall some
elementary notions and conclusions of Boolean algebras and the
orthomodular lattices.

\vskip0.1in

Let $(L, \leq)$ be a partially ordered set.  If for any $a, b\in L$,
its infimum $a\wedge b$ and supremum $a\vee b$ exist, then $(L,
\leq)$ is said to be a {\it lattice}.  If $(L, \leq)$ is a lattice
and for any $a, b, c\in L$, we have
\begin{eqnarray}a\wedge (b \vee c) = (a\wedge b) \vee (a \wedge c), \\a\vee (b\wedge c)=(a\vee b)\wedge(a\vee
c), \end{eqnarray} then we say that $(L,  \leq)$ satisfies the {\it
distributive law}.  Let $(L,  \leq)$ be a lattice with the largest
element $I$ and the smallest element $\theta$. If there exists a
mapping $': L\rightarrow L$ such that for each $a\in L$, $a\vee
a'=I, a\wedge a'=\theta$, $(a')'=a$ and $b'\leq a'$ whenever $a\leq
b$, then $(L,  \leq)$ is said to be an {\it orthogonal complement
lattice}. Let $(L,  \leq)$ be an orthogonal complement lattice. If
$a, b\in L$ and $a\leq b$, we have
$$b=a\vee (b\wedge a'),\eqno (3)$$ then we say that $(L, \leq)$
satisfies the {\it orthomodular law}.

\vskip0.1in

{\bf Definition 2.1} Let $(L,  \leq)$ be an orthogonal complement
lattice. If $(L,  \leq)$ satisfies the distributive law, then $(L,
\leq)$ is said to be a Boolean algebra; if $(L,  \leq)$ satisfies
the orthomodular law, then $(L, \leq)$ is said to be an orthomodular
lattice.

\vskip0.1 in

{\bf Example 2.1} Let $X$ be a set and $2^{X}$ be its all subsets.
Then $(2^{X}, \subseteq)$ is a Boolean algebra.

\vskip0.1 in

{\bf Example 2.2}([4-5]) Let $H$ be a complex Hilbert space, $P(H)$
be the set of all orthogonal projection operators on $H$, $P_1,
P_2\in P(H)$. If we define $P_1\leq P_2$ if and only if
$P_1P_2=P_2P_1=P_1$, then $(P(H), \leq)$ is an orthomodular lattice.

Example 2.2 is the most important and famous quantum logic model
which was introduced in 1936 by Birkhoff and von Neumann ([4]).

\vskip0.1 in

Let $(L, \leq)$ be an orthomodular lattice and $a, b\in L$. If $a
\leq b'$, then we say that $a$ and $b$ are orthogonal and denoted by
$a\perp b$. A subset $\{a_{1},a_{2},\cdots,a_{n} \}$ of $L$ is said
to be an orthogonal set if $a_{1}\perp a_{2},
 (a_{1}\vee a_{2})\perp a_{3},\cdots, (a_{1}\vee a_{2}\vee \cdots a_{n-1})\bot
 a_{n}.$

\vskip0.1 in

Let $(L, \leq)$ be an orthomodular lattice and $s: L\rightarrow [0,
1]$ be a mapping from $L$ into the real number interval $[0,1]$. If
$s(I)=1$ and $s(a\vee b)=s(a)+s(b)$ whenever $a\bot b$, then $s$ is
said to be a state of $(L, \leq)$.

It is clear that if $s$ is a state of $(L, \leq)$ and
$\{a_{1},a_{2},\cdots,a_{n} \}$ is a finite orthogonal subset of
$L$, then $s(\vee_{i=1}^na_i)=\sum_{i=1}^ns(a_i)$.

\vskip0.1 in

In [2], the author defined the following three concepts:

\vskip0.1 in

Let $(L, \leq)$ be an orthomodular lattice and $s$ be a state of
$(L, \leq)$, $\{a_{1},a_{2},\cdots,a_{n} \}$ be a finite orthogonal
subset of $L$. If $s(\vee_{i=1}^{n}a_{i})=1$, then
$\{a_{1},a_{2},\cdots,a_{n} \}$ is said to be a partition of $(L,
\leq)$ with respect to the state $s$. If $\{a_{1},a_{2},\cdots,a_{n}
\}$ is a partition of $(L, \leq)$ with respect to the state $s$ and
for each $b\in L$, $$s(b)=\sum_{i=1}^ns(a\wedge b),$$ then $s$ is
said to have the {\it Bayes property}. Moreover, let
$\{a_{1},a_{2},\cdots,a_{n} \}$ and $\{b_{1},b_{2},\cdots,b_{m} \}$
be two partitions of $(L, \leq)$ with respect to the state $s$. Then
the set $\{a_{i}\wedge b_j: i=1, 2, \cdots, n; j=1, 2, \cdots, m\}$
is said to be a refinement of the partitions
$\{a_{1},a_{2},\cdots,a_{n} \}$ and $\{b_{1},b_{2},\cdots,b_{m} \}$
.

\vskip0.1 in

By the distributive law of Boolean algebra, it is clear that each
state on the Boolean algebra has the Bayes property. However, the
following example shows that there is no state $s$ on $(P(H), \leq)$
with the Bayes property, where $H$ is a complex Hilbert space with
$\dim(H)=2$. Moreover, our example shows also that the concept of
refinement of partitions is also not effective for $(P(H), \leq)$.

\vskip0.1 in

{\bf Example 2.3}  Let $H$ be a complex Hilbert space with
$\dim(H)=2$ and $a_1=\{(0, z): z\in \mathbb{C}\}$, $a_2=\{(z, 0):
z\in \mathbb{C}\}$. If $P_1, P_2$ are the orthogonal projection
operators from $H$ onto $a_1$ and $a_2$, respectively, then for any
state $s$, $A=\{P_1, P_2\}$ is a partition of $(P(H), \leq)$ with
respect to state $s$. Let $b_1=\{(\frac{\sqrt{2}z}{2},
\frac{\sqrt{2}z}{2}): z\in \mathbb{C}\},
b_2=\{(-\frac{\sqrt{2}z}{2}, \frac{\sqrt{2}z}{2}): z\in
\mathbb{C}\}$ and  $Q_1, Q_2$ are the orthogonal projection
operators on $b_1$ and $b_2$, respectively. Then $P_{i}\wedge Q_j=0,
i, j=1, 2.$ So $\vee_{i=1}^{n}(P_{i}\wedge Q_j))=0, j=1, 2.$ If
state $s$ has the Bayes property, then we have $s(Q_j)=s(0)=0, j=1,
2$, so $s(Q_1)+s(Q_2)=0$. On the other hand, note that $Q_1\bot Q_2$
and $Q_1\vee Q_2=I$, so $1=s(I)=s(Q_1)+s(Q_2)=0$, this is a
contradiction and so there is no state $s$ on $(P(H), \leq)$ which
has the Bayes property. Moreover, since $P_{i}\wedge Q_j=0, i, j=1,
2$, so $\{P_i\wedge Q_j: i, j=1, 2\}$ cannot be considered as a
refinement of two partitions $\{P_1, P_2\}$ and $\{Q_1, Q_2\}$.

\vskip0.1 in

Example 2.3 told us that we must redefine the refinement concept of
partitions of quantum logics.

\vskip0.1 in

In quantum theory, we have known that each orthogonal projection
operator can be looked as the {\it sharp measurement}. For two sharp
measurements $P$ and $Q$, if $P$ is performed first and $Q$ second,
then $PQP$ have important physics meaning ([6-8]). If $\{P_{1},
P_{2}, \cdots, P_{n}\}$ and $\{Q_{1}, Q_{2}, \cdots, Q_{m}\}$ are
two orthogonal sets of $(P(H), \leq)$ and $\vee_{i=1}^nP_{i}=I$,
$\vee_{i=1}^mQ_{i}=I$, then we may try to use $$\{Q_{j}P_{i}Q_{j},
i=1, 2, \cdots, n, i=1, 2, \cdots, n; j=1, 2, \cdots, m\}$$ as the
refinement of $\{P_{1}, P_{2}, \cdots, P_{n}\}$ and $\{Q_{1}, Q_{2},
\cdots, Q_{m}\}$. However, note that, in general, $Q_jP_iQ_j$ is not
an orthogonal projection operator on $H$, that is, $Q_jP_iQ_j\notin
P(H)$, so we must to transfer the sharp measurements to unsharp
measurements. In 1994, Foulis and Bennett completed the famous
transformation, that is, they introduced the following algebra
structure and called it as the {\it effect algebra} ([9]):

\vskip0.1 in

Let $(E, \theta, I,  \oplus)$ be an algebra system, where $\theta$
and $I$ be two distinct elements of $E$, $\oplus$ be a partial
binary operation on $E$ satisfying that:

(EA1) If $a\oplus b$ is defined,  then $b\oplus a$ is defined and
$b\oplus a=a\oplus b$.

(EA2)  If $a\oplus (b\oplus c)$ is defined,  then $(a\oplus b)\oplus
c$ is defined and $$(a\oplus b)\oplus c=a\oplus (b\oplus c). $$

(EA3) For every $a\in E$,  there exists a unique element $b\in E$
such that $a\oplus b=I$.

(EA4) If $a\oplus I$ is defined,  then $a=\theta$.

\vskip0.1 in

In an effect algebra $(E, \theta, I,  \oplus)$, if $a\oplus b$ is
defined, we write $a\bot b$. For each $a\in E$, it follows from
(EA3) that there exists a unique element $b\in E$ such that $a\oplus
b=1$, we denote $b$ by $a'$. Let $a, b\in E$, if there exists an
element $c\in E$ such that $a\bot c$ and $a\oplus c=b$, then we say
that $a\leq b$. It follows from [9] that $\leq $ is a partial order
of $(E,0,1, \oplus)$ and satisfies that for each $a\in E$, $0\leq
a\leq 1$, $a\bot b$ if and only if $a\leq b'$. If $a\wedge a'=0$,
then $a$ is said to be a {\it sharp element} of $E$.

\vskip0.1 in

Let $H$ be a complex Hilbert space. A self-adjoint operator $A$ on
$H$ such that $0\leq A\leq I$ is called a {\it quantum effect} on
$H$ ([6-9]). If a quantum effect represents a measurement, then the
measurement may be {\it unsharp} ([6, 9]). The set of quantum
effects on $H$ is denoted by $E(H)$. For $A, B\in E(H)$, if we
define $A\oplus B$ if and only if $A+B\leq I$ and let $A\oplus
B=A+B$, then $(E(H), \theta, I, \oplus)$ is an effect algebra, and
its all sharp elements are just $P(H)$ ([5-6, 9]).

\vskip0.1 in

Moreover, Professor Gudder introduced and studied the following {\it
sequential effect algebra} theory ([10-11]):

Let $(E, \theta, I, \oplus)$ be an effect algebra and another binary
operation $\circ $ defined on $(E,\theta, I, \oplus)$ satisfying
that

(SEA1) The map $b\mapsto a\circ b$ is additive for each $a\in E$,
that is, if $b\bot c$, then $a\circ b\bot a\circ c$ and $a\circ
(b\oplus c)=a\circ b\oplus a\circ c$.

(SEA2) $I\circ a=a$ for each $a\in E$.

(SEA3) If $a\circ b=\theta$, then $a\circ b=b\circ a$.

(SEA4) If $a\circ b=b\circ a$, then $a\circ b'=b'\circ a$ and for
each $c\in E$, $a\circ (b\circ c)=(a\circ b)\circ c$.

(SEA5) If $c\circ a=a\circ c$ and $c\circ b=b\circ c$, then
$c\circ(a\circ b)=(a\circ b)\circ c$ and $c\circ(a\oplus b)=(a\oplus
b)\circ c$ whenever $a\bot b$.

\vskip 0.1 in

Let $(E,\theta, I, \oplus, \circ)$ be a sequential effect algebra.
If $a, b\in E$ and $a\circ b=b\circ a$, then we say that $a$ and $b$
is {\it sequentially independent} and denoted by $a|b$.

Now, we use the sequential effect algebra theory as tools to study
the partitions and refinements of quantum logics and their
entropies.

\vskip 0.2 in

\noindent{\bf 3. Partitions, refinements and their entropies}

\vskip0.2in

Let $(E,\theta, I, \oplus, \circ)$ be a sequential effect algebra. A
set $\{a_{1},  a_{2},  \cdots, a_{n}\}$ is said to be a partition of
$(E, \theta, I,  \oplus, \circ)$ if $\oplus_{i=1}^na_i$ is defined
and $\oplus_{i=1}^{n} a_{i}=I$.

\vskip 0.1 in

In following, we denote partitions $A=\{a_{1},
 a_{2},  \cdots,  a_{n}\}$, $B=\{b_{1},  b_{2},  \cdots,  b_{m}\}$,
 $C=\{c_{1},  c_{2},  \cdots,  c_{l}\}$, and $A\circ
B=\{a_{i}\circ b_{j}: a_{i}\in A, b_{j}\in B, i=1, 2, \cdots, n,
j=1, 2, \cdots, m\}$. That $A\circ B\neq B\circ A$ are clear.

\vskip0.1in

Let $(E, \theta, I,  \oplus, \circ)$ be a sequential effect algebra,
$A$ and $B$ be two partitions of $(E, \theta, I,  \oplus, \circ)$.
Then it follows from (SEA1) and ([11, Lemma 3.1(i)]) that $A\circ B=
\{a_{i}\circ b_{j}: a_{i}\in A, b_{j}\in B, i=1, 2, \cdots, n, j=1,
2, \cdots, m\}$ is also a partition of $(E,\theta, I, \oplus,
\circ)$. We say that the partition $A\circ B$ is a refinement of the
partitions $A$ and $B$.

\vskip 0.1 in

{\bf Example 3.1}([11]) Let $(L, \leq)$ be a Boolean algebra, $a,
b\in L$. Let $a\oplus b$ be defined iff $a\wedge b=\theta$, in this
case, $a\oplus b=a\vee b$, and define $a\circ b=a\wedge b$. Then
$(L,\theta, I, \oplus, \circ)$ is a sequential effect algebra.

\vskip 0.1 in

{\bf Example 3.2}([11]) Let $X$ be a set and $\mathcal{F}$$(X)$ be
the all fuzzy sets of $X$, $\mu_{\tilde{A}}, \mu_{\tilde{B}}\in
\mathcal{F}$$(X)$. Let $\mu_{\tilde{A}}\oplus \mu_{\tilde{B}}$ be
defined iff $\mu_{\tilde{A}}+\mu_{\tilde{B}}\leq 1$, in this case,
$\mu_{\tilde{A}}\oplus
\mu_{\tilde{B}}=\mu_{\tilde{A}}+\mu_{\tilde{B}}$, and define
$\mu_{\tilde{A}}\circ \mu_{\tilde{B}}=\mu_{\tilde{A}}
\mu_{\tilde{B}}$. Then $(\mathcal{F}$$(X), 0, 1, \oplus, \circ)$ is
a sequential effect algebra.

\vskip 0.1 in

{\bf Example 3.3}([11]) Let $H$ be a complex Hilbert space, if for
any two quantum effects $B$ and $C$, we define $B\circ
C=B^{\frac{1}{2}}CB^{\frac{1}{2}}$, then $({\cal E}(H), 0, I,
\oplus, \circ)$ is a sequential effect algebra. In particular, for
any two orthogonal projection operators $P$ and $Q$ on $H$,
$PQP=P^{\frac{1}{2}}QP^{\frac{1}{2}}$ is a sequential product of $P$
and $Q$.

\vskip0.1in

The above three examples showed that our refinement methods of the
partitions are not only suitable for classical logics, but also
effective for fuzzy logics and quantum logics.

\vskip0.1in

Now, we begin to study the entropies of partitions and refinements
of sequential effect algebras. First, we need the following:

\vskip0.1in

Let $(E, \theta, I,  \oplus, \circ)$ be a sequential effect algebra,
$s$ be a state of $(E, 0, 1, \oplus, \circ)$, that is, $s:
E\rightarrow [0, 1]$ be a mapping from $E$ into the real number
interval $[0,1]$ such that $s(I)=1$ and whenever $a\oplus b$ be
defined, $s(a\oplus b)=s(a)+s(b)$. Then for given $A$, $$s_A:
b\rightarrow \sum_{i=1}^{n}s(a_{i}\circ b)$$ defines a new state
$s_{A}$, this is the resulting state after the system $A$ is
executed but no observation is performed ([12]). Moreover, we denote
$s(b\mid a)$ by $s(a\circ b)/s(a)$ if $s(a)\neq 0$ and $0$ if
$s(a)=0$.

\vskip0.1in

The entropy of $A$ with respect to the state $s$ is defined by
$$H_{s}(A)=-\sum_{i=1}^{n}s(a_{i})\log s(a_{i}).$$

\vskip0.1in

The {\it refinement entropy} of $A$ and $B$ with respect to the
state $s$ is defined by
$$H_{s}(A\circ B)=-\sum_{i=1}^{n}\sum_{j=1}^{m}s(a_{i}\circ
b_{j})\log s(a_{i}\circ b_{j}). $$

\vskip0.1in

The {\it conditional entropy} of $B$ conditioned by $A$ with respect
to the state $s$ is defined by
$$H_{s}(B|
A)=-\sum_{i=1}^{n}\sum_{j=1}^{m}s(a_i\circ b_{j})\log s(b_{j}|
a_{i}). $$

\vskip0.1in

{\bf Lemma 3.1}([13]) ({\it log sum inequality}) For non-negative
numbers $a_{1}, a_{2},\cdots, a_{n}$ and $b_{1}, b_{2}, \cdots,
b_{n}$,
\begin{eqnarray*}
\sum_{i=1}^{n}a_{i}\log \frac{a_{i}}{b_{i}}\geq
(\sum_{i=1}^{n}a_{i})\log(\frac{\sum_{i=1}^{n}a_{i}}{\sum_{i=1}^{n}b_{i}}).
\end{eqnarray*}

\vskip0.1in

We use the convention that $0\log 0 = 0, a\log \frac{a}{0}=\infty$
if $a
>0$ and $0\log\frac{0}{0}=0.$

\vskip0.1in

In this paper, our main result is the following theorem which
generalizes the classical entropy properties ([2, 13-14]) to the
sequential effect algebras.

\vskip0.1in

{\bf Theorem 3.1} (i). $H_{s}(A\circ B)= H_{s}(B| A)+
 H_{s}(A)$.

(ii). $H_{s}(A|C)\leq H_{s}(A\circ B|C)$.

(iii). $H_{s}(B| A)\leq H_{s_A}(B)$.

(iv). $H_{s}(A\circ B)\leq H_{s}(A)+ H_{s_A}(B)$.

(v). $\max \{H_{s_{A}}(B), H_{s}(A)\}\leq H_{s}(A\circ B)$.

(vi). $H_{s}(B\circ A|C)\leq H_{s_C}(A|B)+H_{s}(B|C)$.

\vskip0.1in
 {\bf Proof}. We only prove (vi). In fact, by Lemma 3.1, we have
\begin{eqnarray*}
&\quad&H_{s_C}(A|B)+H_{s}(B|
C)\\&=&-\sum_{i=1}^{n}\sum_{j=1}^{m}s_C(b_{j}\circ a_{i})\log
s_C(a_{i}|b_{j})-\sum_{j=1}^{m}\sum_{k=1}^{l}s(c_{k}\circ b_{j})\log
s(b_{j}|c_{k})
\\&=&-\sum_{i=1}^{n}\sum_{j=1}^{m}\sum_{k=1}^{l}s(c_{k}\circ (b_{j}\circ a_{i}))\log \frac{\sum_{k=1}^{l}s(c_{k}\circ (b_{j}\circ a_{i}))}{\sum_{k=1}^{l}s(c_{k}\circ b_{j})}
\\&\quad&-\sum_{j=1}^{m}\sum_{k=1}^{l}s(c_{k}\circ b_{j})\log
\frac{s(c_{k}\circ b_{j})}{s(c_{k})}\\&\geq
&-\sum_{i=1}^{n}\sum_{j=1}^{m}\sum_{k=1}^{l}s(c_{k}\circ (b_{j}\circ
a_{i}))\log \frac{s(c_{k}\circ (b_{j}\circ a_{i}))}{s(c_{k}\circ
b_{j})}
\\&\quad&-\sum_{j=1}^{m}\sum_{k=1}^{l}s(c_{k}\circ b_{j})\log
\frac{s(c_{k}\circ
b_{j})}{s(c_{k})}\\&=&-\sum_{i=1}^{n}\sum_{j=1}^{m}\sum_{k=1}^{l}
s(c_{k}\circ (b_{j}\circ a_{i}))\log \frac{s(c_{k}\circ (b_{j}\circ
a_{i}))}{s(c_{k})}
\\&=&H_{s}(B\circ A|C).
\end{eqnarray*}
 That concludes the proof.

 \vskip0.1in

Finally, we would like to point out that for the advances of
sequential effect algebras, see [15-18].

\vskip0.2in

\centerline{\bf References}

 \vskip0.2in

\noindent [1] Ohya M.  and  Petz D. {\it Quantum Entropy and its
Use. } Springer-Verlag,  Berlin,  (1991)

\noindent [2] Yuan Hejun.  J. Entropy of Partitions on Quantum
Logic. Commun. Theor. Phys. {\bf 43},  437-439,  (2005)

\noindent [3]  Kelly J.  L. {\it General Topology. }
Springer-Verlag, New York,  (1955)

\noindent [4] Birkhoff G. and  von Neumann J. The logic of quantum
mechanics, Ann. Math. {\bf 37}, 823-834, (1936)

\noindent [5] Dvure$\check{c}$enskij A. and Pulmannov$\check{a}$ S.
{\it New Trends in Quantum Structures.} Kluwer,  (2002)

\noindent [6] Busch P.,  Grabowski M.  and  Lahti P.  J. {\it
Operational Quantum Physics.}  Springer-Verlag,  Berlin,  (1995)

\noindent [7] Busch P.,  Lahti P.  J.  and Mittlestaedt P. {\it The
Quantum Theory of Measurements.}  Springer-Verlag,  Berlin, (1991)

\noindent [8] Busch P.  and  Singh J. L\"{u}ders theorem for unsharp
quantum measurements.  Phys.  Lett.  A. {\bf 249},  10-12, (1998)

\noindent [9] Foulis D.  J. and Bennett M. K. Effect algebras and
unsharp quantum logics. Found. Phys.  {\bf 24}, 1331-1352, (1994)

\noindent [10] Gudder S.   and  Nagy G. Sequential quantum
measurements.  J.  Math.  Phys.  {\bf 42},  5212-5222,  (2001)

\noindent [11] Gudder S.  and  Greechie R. Sequential products on
effect algebras.  Rep.  Math.  Phys.  {\bf 49},  87-111,  (2002)

\noindent [12] Arias A., Gheondea A.  and  Gudder S. Fixed points of
quantum operations.  J.  Math. Phys. {\bf 43}, 5872-5881, (2002)

\noindent [13] Cover T.  M.  and Thomas J. A. {\it Elements of
Information Theory. } Wiley,  New York, (1991)

\noindent [14] Zhao Yuexu and Ma Zhihao. Conditional Entropy of
Partitions on Quantum Logic.  Commun.  Theor.  Phys.  {\bf 48},
11-13 (2007)

\noindent [15] Shen Jun and Wu Junde. Not each sequential effect
algebra is sharply dominating. Phys. Letter A. {\bf 373}, 1708-1712,
(2009)

\noindent [16] Shen Jun and Wu Junde. Remarks on the sequential
effect algebras. Report. Math. Phys. {\bf 63}, 441-446, (2009)

\noindent [17] Liu Weihua and Wu Junde. A uniqueness problem of the
sequence product on operator effect algebra ${\cal E}(H)$. J. Phys.
A: Math. Theor. {\bf 42}, 185206-185215, (2009)

\noindent [18] Shen Jun and Wu Junde. Sequential product on standard
effect algebra ${\cal E}(H)$. J. Phys. A: Math. Theor. {\bf 44},
(2009)

\end{document}